 \documentclass[doublecol]{epl2}

\usepackage{amssymb}

\usepackage{epstopdf}
\usepackage{color}

\title{Strong enhancement of third harmonic generation
 in a double layer graphene system caused by electron-hole pairing}
\shorttitle{Enhancement of third harmonic generation caused by electron-hole pairing} 

\author{K. V. Germash\inst{1} \and D. V. Fil\inst{1,2}}
\shortauthor{K. V. Germash and D. V. Fil}

\institute{
  \inst{1} Institute for Single Crystals, National Academy of Sciences of Ukraine,
  Nauki ave. 60, Kharkov 61001, Ukraine
\\
  \inst{2} V. N. Karazin Kharkov National University, Svobody Sq. 4, Kharkov 61022, Ukraine
}


\abstract{ A manifestation of electron-hole pairing in nonlinear electromagnetic response
of a double layer graphene system is studied. It is shown that the pairing causes the
appearance of a number of peaks in the frequency dependence of the intensity of the third
harmonic generation (THG).  The highest peak corresponds to $\hbar\omega = (2/3)\Delta$,
where $\omega$ is the incident wave frequency, and $\Delta$ is the order parameter of the
electron-hole pairing. The absolute value of the THG intensity in the systems with
electron-hole pairing is in several orders of magnitude greater than the THG intensity in
the unpaired state. It is shown that huge enhancement of the THG intensity occurs both in
the double monolayer and double bilayer graphene systems. }

\begin{document}

\maketitle

\section{Introduction}

Nonlinear optical and microwave properties of graphene attract considerable attention.
Strong nonlinear response of graphene can be seen from the classical equation of motion
for a charged particle with the spectrum linear in the momentum\cite{1,2}. This equation
yields the electric current that contains all odd Fourier harmonics with the amplitudes
falling down very slowly with the harmonic number. The quantum approach \cite{3,4,5,5a,6}
predicts resonant behavior for the  {third harmonic generation} in graphene. The
frequency dependence of the THG intensity has the main peak at $\hbar\omega =
(2/3)\varepsilon_F$ and two minor peaks at $\hbar\omega = \varepsilon_F$ and $\hbar\omega
= 2\varepsilon_F$, where $\varepsilon_F$ is the Fermi energy. Nonlinear optical effects
in graphene systems were observed experimentally. In \cite{7}  the coherent nonlinear
optical response of single- and few-layer graphene was measured using four-wave mixing.
Sharp contrast images of graphene flakes on a dielectric substrate at combined
frequencies were observed. Graphene with the effective thickness 3 $\AA$ demonstrates the
nonlinear emission intensity  in 10 times larger than a 4 nm gold film. Nonlinear
susceptibility at the wavelength $\lambda\approx 1$ $\mu$m is evaluated as large as
$10^{-7}$ esu. In the THG experiment \cite{8} the effective nonlinear susceptibility
$\chi^{(3)}\sim 5\cdot 10^{-9}$ esu at the incident wavelength $\lambda_i=1.7$ $\mu$m was
registered. Strong THG in a monolayer graphene on an amorphous silica substrate was
reported in \cite{9}. It was shown that the effective nonlinear susceptibility of
graphene is in 4.6 times larger than of a thick Au film  (for $\lambda_i=789$ nm).

It is known that linear optical properties of graphene are also quite unusual. In a wide
frequency range the absorption coefficient $A$ of  pristine graphene is determined by the
fundamental constants: $A=\pi e^2/\hbar c$ \cite{10}. In a doped graphene  the absorption
is suppressed in the frequency range $\hbar\omega<2 \varepsilon_F$ due to the reduction
of the interband transitions.

Graphene  double layer systems are considered as  perspective ones for achieving the
electron-hole pairing. The electron-hole pairing is an analog of the Cooper pairing. It
may reveal itself in a so-called  counterflow superconductivity. The effect can also be
understood as the superfluidity of a gas of electron-hole pairs. The possibility of
electron-hole pairing in a double monolayer graphene (DMLG) system was considered in
 {\cite{22,21a,21,23}}. The pairing in DMLG in the quantum Hall regime was analyzed in
\cite{24,25,26}. In \cite{11} we have considered electromagnetic properties of  DMLG in
the paired state and find resonant behavior of the absorption and reflection coefficients
at the photon energy equal to the excitonic energy gap in the spectrum
($\hbar\omega=2\Delta$).

The general idea of the electron-hole pairing in double layer systems was put forward in
\cite{12,13}  well before the proposals \cite{22,21a,21,23}.  The electron-hole pairing
in quantum Hall systems was predicted in \cite{14,15,16}. The pairing was registered in
double quantum well AlGaAs heterostructures in the quantum Hall state under study of
their transport properties \cite{17,18,19,20}.
 The possibility of electron-hole pairing  was
also considered with reference to topological insulator heterostructures
\cite{t1,t2,t3,t4}, double bilayer graphene \cite{27},  double few-layer graphene
\cite{28}, transition metal dichalcogenide \cite{tm1,tm2,tm3} and black phosphorene
\cite{bp1} double layers. Recently several experimental efforts to register electron-hole
pairing in  {double layer graphene systems} were done \cite{ge1,ge2,ge3,ge4}, but the
results of these experiments are controversial.

 The electron-hole
pairing is caused by the Coulomb attraction. The bare Coulomb interaction is a strong
one, but screening reduces this interaction that may result in a lowering of the critical
(superfluid transition) temperature down to the $\mu$K range \cite{29}. At the same time
the pairing suppresses the screening \cite{my3}. Mutual influence of pairing and
screening was analyzed in \cite{30,31}. It was found that the behavior of the system
 {is very} sensitive to the value of the dimensionless parameter $r_s$, the ratio of the
average Coulomb interaction energy to the Fermi energy. For the monolayer graphene this
parameter is a material constant independent of the density of carriers: $r_s=e^2/\hbar
v_F \varepsilon_{eff}$, where $v_F$ is the graphene Fermi velocity, and
$\varepsilon_{eff}$ is the effective dielectric constant (for a double-layer system on a
dielectric substrate $\varepsilon_{eff}=(\varepsilon+1)/2$, where $\varepsilon$ is the
substrate dielectric constant). According to \cite{30}, the screening is suppressed and
the critical
 temperature   reaches $T_c\sim 0.1 \varepsilon_F$, if the
parameter $r_s$ exceeds the critical value $r_s^{(c)}\approx 1.5$. This condition is
fulfilled at $\varepsilon_{eff}\lesssim 1.5$. It may correspond, for instance, a porous
dielectric substrate with $\varepsilon\approx 2$ \cite{p2}. For the bilayer graphene  the
parameter $r_s$ depends on the density of carriers: $r_s=(a_B^* k_F)^{-1}$, where $ k_F$
is the Fermi wave number, $a_B^*=\hbar^2 \varepsilon_{eff} /m_{eff} e^2$ is the effective
Bohr radius and $m_{eff}$ is the electron  effective mass. In double bilayer graphene
(DBLG) systems the screening is suppressed at $r_s>r_s^{(c)}\approx 5$, and since $r_s$
increases under decrease in $k_F$, the latter condition can be achieved for crystalline
dielectric substrates ($\varepsilon\approx 4$) as well.

\section{Intensity of THG in a double layer system}

 The
system under study consists of two monolayer or two bilayer graphene sheets (layers)
separated by a thin dielectric layer. We imply that the concentration of conducting
electrons in one layer is equal to the concentration of holes in the other layer. The
incident wave with the frequency $\omega$ induces high frequency electric currents in the
graphene layers. The component of the current that oscillates with the frequency
$3\omega$  is responsible for THG. Taking into account the electron-hole symmetry one can
write   nonlinear in the electric field current as
\begin{eqnarray} \label{1}
  j_+^{(3)}&=& \sigma^{(3)}_{+++} E_{+}^3
  +\sigma^{(3)}_{+--} E_{+} E_{-}^2,
  \cr
  j_-^{(3)}&=& \sigma^{(3)}_{---} E_-^3+\sigma^{(3)}_{-++}E_- E_+^2,
\end{eqnarray}
where $j_\pm^{(3)}=j_1^{(3)}\pm j_2^{(3)}$, $E_{\pm}= E_1\pm E_2$, $j_i$ is the electric
current in the layer $i$ and $E_i$ is the electric field in this layer. Eqs. (\ref{1})
are presented in the symbolic form. In the general case they are matrix integral
equations that account tensor nature of the conductivity and nonlocality of the response
in time and space. In the case of uncoupled layers (unpaired electrons and holes) the
conductivities in (\ref{1}) are expressed through $\sigma_1^{(3)}$, the 3rd order
single-layer conductivity : $\sigma^{(3)}_{+++}=\sigma^{(3)}_{---}=\sigma^{(3)}_{1}/4$,
$\sigma^{(3)}_{+--}=\sigma^{(3)}_{-++}=3\sigma^{(3)}_{1}/4$.

 {In this study we put $E_-=0$ due to the following reason. The ratio of
amplitudes of the fields $E_-$ and $E_+$ is evaluated as $|E_-|/|E_+|\approx q_z d/2$,
where $q_z$ is the normal component of the wave vector of the incident wave, and $d$ is
the interlayer distance. We consider the frequency range $\hbar \omega\lesssim
\varepsilon_F$. In this range $|E_-|/|E_+|\lesssim k_F d v_F/c$, where $c$ is the
velocity of light. The electron-hole pairing occurs at rather small interlayer distance
$d$. In particular, for the double monolayer graphene the condition $d k_F\lesssim 0.1$
 should be fulfilled \cite{30}. Therefore
$|E_-|/|E_+|< 10^{-3}$ and nonlinear response to the small field $E_-$ can be neglected.}

 Let  the electric component of the incident wave is
$E_x(z,t)=E_0 \cos (q_z z-\omega t)$. Then, the 3rd harmonic of the electric current  is
given by the expression
 \begin{equation}\label{2}
    j^{(3\omega)}_{x,+}(t)=\sigma_{+++}^{(3)}(\omega,\omega,\omega) E_{0}^3
    e^{- 3 i\omega t}+c.c.,
\end{equation}
where $\sigma^{(3)}_{ {+++}}(\omega,\omega,\omega) $   {is} the $xxxx$ component of the
high-frequency nonlinear conductivity tensor.

The boundary conditions determine the relation between the current (\ref{2}) and the
magnetic component of the generated wave. It yields
\begin{equation}\label{3}
B_{y}^{(3)}(z,t)=\mp\frac{2\pi}{c}\sigma_{+++}^{(3)}(\omega,\omega,\omega) E_{0}^3 e^{\pm
3i q_z z- 3i\omega t}+c.c.,
\end{equation}
where sign $\pm$ corresponds to different half-spaces.  {The boundary condition (\ref{3})
assumes that the double layer system is considered as a zero-thickness conducting layer.
It is equivalent to the limit $q_z d \to 0$.}

The intensity of the third harmonic is given by the Poynting vector averaged over the
period $T$:
\begin{equation}\label{3a}
   I^{(3)}=\frac{c}{4\pi T}\int_0^T
|\mathbf{E}^{(3)}\times \mathbf{B}^{(3)} | dt.
\end{equation}
 Assuming, for simplicity, that the
dielectric constant of the environment  is $\varepsilon=1$,  we
obtain
\begin{equation}\label{4}
I^{(3)}=4\left(\frac{2
\pi}{c}\right)^4|4\sigma_{+++}^{(3)}(\omega,\omega,\omega)|^2I_{inc}^3,
\end{equation}
where $I_{inc}$ is the intensity of the incident wave.  For two uncoupled layers
($\sigma_{+++}^{(3)}=\sigma_{1}^{(3)}/4$) Eq. (\ref{4}) corresponds to the quadruplicate
intensity of a single layer. The factor 4 is due to the constructive interference.

\section{Electron-hole pairing in the double layer graphene}

In what follows we use the Dirac Hamiltonian that describes electromagnetic properties of
graphene in the low-energy approximation. In this approximation two valleys near Dirac
points $K$ and $K'$ are considered independently and each valley has two spin components.
Each Dirac component yields the same contribution into the nonlinear conductivity. Below
we consider only one component and take into account the other ones by the factor 4 in
the final answer.

The Hamiltonian of the DMLG system has the form
\begin{eqnarray}\label{5}
{H} = \sum_{i,\mathbf{k},\lambda} \xi_{i\mathbf{k}\lambda} {c}^{+}_{i,\mathbf{k},\lambda}
{c}_{i,\mathbf{k},\lambda}
 +\frac{1}{2S} \sum_{i,j,\mathbf{q}}
V_{ij}(q) : \hat{n}_{i,\mathbf{q}} \hat{n}_{j,-\mathbf{q}}: \cr +
 \frac{1}{S}
\sum_{i,\mathbf{q}}e\varphi_{i,\mathbf{q}}(t) \hat{n}_{i,-\mathbf{q}},
\end{eqnarray}
where ${c}^{+}_{i,\mathbf{k},\lambda} ({c}_{i,\mathbf{k},\lambda} )$ is the creation
(annihilation) operator for the electron in the layer $i$ in the state with the momentum
$\mathbf{k}$ and the subband index $\lambda=\pm 1$. This state corresponds to the energy
$\xi_{i\mathbf{k}\lambda}=\varepsilon_{\mathbf{k}\lambda}-\mu_i$, where
$\varepsilon_{\mathbf{k}\lambda}=\lambda\hbar v_F k$ is the electron spectrum of the
monolayer graphene near  $K$ and $K'$ point, and $\mu_i$ is the chemical potential in the
layer $i$. In what follows we imply $\mu_1=-\mu_2=\mu$ and neglect the difference between
$\mu$ and $\epsilon_F$. In the Hamiltonian (\ref{5}) $\hat{n}_{i,\mathbf{q}}$ is the
electron density operator, $V_{ij}(q)$ is the Fourier
 component of the Coulomb interaction  {energy between electrons in the layers i and j}, $\varphi_{i,\mathbf{q}}(t)$ is the Fourier component of the
  scalar potential  {of the external electromagnetic field}, the
notation $: \hat{O}:$ indicates the normal ordering of creation and annihilation
operators, and $S$ is the area of the system. The explicit expression for the electron
density operator reads
\begin{equation}\label{6}
\hat{n}_{i,\mathbf{q}} = \sum_{\mathbf{k},\lambda,\lambda'}
g_{\mathbf{k}+\mathbf{q},\lambda',\mathbf{k},\lambda} {c}^{+}_{i,\mathbf{k} +
\mathbf{q},\lambda'} {c}_{i,\mathbf{k},\lambda},
\end{equation}
where
\begin{equation}\label{6a}
  g_{\mathbf{k}_1,\lambda_1,\mathbf{k}_2,\lambda_2}= \frac{ e^{\frac{i}{2}( \vartheta_{\mathbf{k}_1} -
\vartheta_{\mathbf{k}_2} )} + \lambda_1 \lambda_2 e^{-\frac{i}{2}(
\vartheta_{\mathbf{k}_1} - \vartheta_{\mathbf{k}_2} )}}{2}
\end{equation}
and $\vartheta_{\mathbf{k}}$ is the angle between $\mathbf{k}$ and the $x$-axis.

In (\ref{5}) we use the gauge in which the vector potential is directed normally to the
layers and the in-plane electric field is given by the scalar potential $E_x=-\partial
\varphi /\partial x$. Such a gauge can be used if the $x$ component ($q_x$) of the wave
vector  of the incident wave is nonzero. The response at normal incidence can be computed
as the limit $q_x\to 0$.

The order parameter for the electron-hole pairing is given by the equation
\begin{eqnarray} \label{7}
\Delta_{\mathbf{k}\lambda} = \frac{1}{S} \sum_{\mathbf{q},\lambda'}
V_{12}(q) \frac{1+\lambda\lambda'\cos(\vartheta_{\mathbf{k} +
\mathbf{q}} - \vartheta_{\mathbf{k}})}{2}\cr \times \overline{
    \langle
    {c}^{+}_{1,\mathbf{k} + \mathbf{q},\lambda'}
    {c}_{2,\mathbf{k} + \mathbf{q},-\lambda'}
    \rangle
}.
\end{eqnarray}
The average in (\ref{7}) is defined as $\overline{\langle \ldots \rangle}=\mathrm{Tr}(
\hat{\rho}\ldots)$, where $\hat{\rho}$ is the density matrix. This average is nonzero in
the state in which the number of electrons in one layer and the number of holes in the
other layer are indefinite, and, in this sense, it can be considered as the anomalous
average.

The mean-field Hamiltonian has the form
\begin{equation}\label{8}
{H}_{MF}(t) =H_0+H_{int}(t),
\end{equation}
where
\begin{eqnarray}\label{9}
{H}_{0} = \sum_{\mathbf{k},\lambda} \bigg[ \xi_{\mathbf{k}\lambda} \bigg(
{c}^{+}_{1,\mathbf{k},\lambda} {c}_{1,\mathbf{k},\lambda} -
{c}^{+}_{2,\mathbf{k},-\lambda} {c}_{2,\mathbf{k},-\lambda} \bigg) \cr -
\left(\Delta_{\mathbf{k}\lambda} {c}^{+}_{2,\mathbf{k},-\lambda}
{c}_{1,\mathbf{k},\lambda} + H.c.\right) \bigg],
\end{eqnarray}
\begin{eqnarray}\label{10}
{H}_{int}(t) = \frac{1}{2S} \sum_{\mathbf{q}} e\varphi_{+,\mathbf{q}}(t)
\hat{n}_{+,-\mathbf{q}},
\end{eqnarray}
$\xi_{\mathbf{k}\lambda}=\varepsilon_{\mathbf{k}\lambda}-\mu$,
$\varphi_+=\varphi_1+\varphi_2$, and $\hat{n}_+=\hat{n}_1+\hat{n}_2$.  {Due to the same
reason as for $E_-$ we put $\varphi_-=\varphi_1-\varphi_2=0$ in Eq. (\ref{10}).}

It is known that in conventional superconductors the order parameter fluctuations
influence the linear and nonlinear response to the external electromagnetic field
\cite{s1,s2}. Nevertheless, in (\ref{10}) we do not account such fluctuations. The reason
is the following.  In the case of the electron-hole pairing the order parameter
fluctuations are coupled with the field $\varphi_-$. The field $\varphi_-$ induces the
fluctuations of the anomalous average in (\ref{7}) that in its turn induces  {the
variation of the difference of electron densities in the layers  $n_-$}. It results in a
renormalization of the linear and nonlinear response to the field $\varphi_-$. On the
contrary, the field $\varphi_+$ is decoupled from the amplitude and phase fluctuations of
the order parameter, and  the response to the field $\varphi_+$ is not modified under
accounting the order parameter fluctuations.

To proceed further we apply  the $u-v$ transformation that diagonalizes the Hamiltonian
${H}_{0}$. The transformation reads
\begin{eqnarray} \label{100}
  {c}_{1,\mathbf{k},\lambda} &=& u_{\mathbf{k}\lambda}{{a}_{\alpha,\mathbf{k},\lambda}}+
  v_{\mathbf{k}\lambda}{{a}_{\beta,\mathbf{k},\lambda}} {,}\cr
  {{c}_{2,\mathbf{k},-\lambda}}  &=&
  u_{\mathbf{k}\lambda}{{a}_{\beta,\mathbf{k},\lambda}}-v_{\mathbf{k}\lambda}
  {{a}_{\alpha,\mathbf{k},\lambda}}
  {,}
\end{eqnarray}
where  ${a}_{\alpha(\beta),\mathbf{k},\lambda}$ are new second quantization operators
that satisfy Fermi anticommutation relations. The
 coefficients in (\ref{100}) are expressed as
\begin{equation}\label{s1}
u_{\mathbf{k}\lambda}=
\sqrt{\frac{1}{2}\left(1+\frac{\xi_{\mathbf{k}\lambda}}{E_{\mathbf{k}\lambda}}\right)},\quad
v_{\mathbf{k}\lambda}=
\sqrt{\frac{1}{2}\left(1-\frac{\xi_{\mathbf{k}\lambda}}{E_{\mathbf{k}\lambda}}\right)},
\end{equation}
 where $E_{\mathbf{k}\lambda}=
\sqrt{\xi^2_{\mathbf{k}\lambda}+\Delta^2_{\mathbf{k}\lambda}}$ are the eigenenergies of
the Hamiltonian $H_0$.

The transformation (\ref{100}) reduces (\ref{9}) and (\ref{10}) to the form
\begin{equation}\label{11}
   {H}_0=\sum_{\nu} E_{\nu}
    {{a}^+_{\nu}{a}_{\nu}},
\end{equation}
\begin{equation}\label{12}
{H}_{int}(t) = \frac{1}{2S}{\sum_{\mathbf{q}}}
\sum_{\nu_1,\nu_2}e\varphi_{+,-\mathbf{q}}(t)
\delta_{\mathbf{k}_2,\mathbf{k}_1-\mathbf{q}} R_{\nu_1,\nu_2}{{a}^+_{\nu_1}}
{{a}_{\nu_2}}.
\end{equation}
Here for the one-particle states  we introduce the notation
$\nu\equiv(\eta,\mathbf{k},\lambda)$, where $\eta=\alpha (\beta)$ corresponds to the
positive (negative) energy $E_{\alpha{,}\mathbf{k},\lambda}= E_{\mathbf{k}\lambda}$
($E_{\beta{,}\mathbf{k},\lambda}=-
 E_{\mathbf{k}\lambda}$). The components of the matrix $\mathbf{R}$
 in (\ref{12}) read
\begin{eqnarray} \label{11a}
  R_{\alpha,\mathbf{k}_1,\lambda_1,\alpha,\mathbf{k}_2,\lambda_2}=
  R_{\beta,\mathbf{k}_1,\lambda_1,\beta,\mathbf{k}_2,\lambda_2}=\cr
  g_{\mathbf{k}_1,\lambda_1,\mathbf{k}_2,\lambda_2}\left(
  u_{\mathbf{k}_1\lambda_1}u_{\mathbf{k}_2\lambda_2}+
  v_{\mathbf{k}_1\lambda_1}v_{\mathbf{k}_2\lambda_2}\right), \cr
 R_{\alpha,\mathbf{k}_1,\lambda_1,\beta,\mathbf{k}_2,\lambda_2}=
  -R_{\beta,\mathbf{k}_1,\lambda_1,\alpha,\mathbf{k}_2,\lambda_2}=\cr
  g_{\mathbf{k}_1,\lambda_1,\mathbf{k}_2,\lambda_2}\left(
  u_{\mathbf{k}_1\lambda_1}v_{\mathbf{k}_2\lambda_2}-
  v_{\mathbf{k}_1\lambda_1}u_{\mathbf{k}_2\lambda_2}\right).
\end{eqnarray}

The DBLG system is treated analogously. The monolayer graphene spectrum is replaced with
the bilayer one:  $\varepsilon^{(b)}_{\mathbf{k}\lambda}=\lambda\hbar^2 k^2/2 m_{eff}$,
and the factor $g$  is modified as
\begin{equation}\label{6ab}
  g^{(b)}_{\mathbf{k}_1,\lambda_1,\mathbf{k}_2,\lambda_2}= \frac{e^{i( \vartheta_{\mathbf{k}_1} - \vartheta_{\mathbf{k}_2} )}
+ \lambda_1 \lambda_2 e^{-{i}( \vartheta_{\mathbf{k}_1} - \vartheta_{\mathbf{k}_2} )}
}{2}.
\end{equation}

 \section{Nonlinear conductivity}

To compute nonlinear conductivity we use the density matrix approach \cite{bo}.  The
density matrix satisfies the  Liouville {-von Neumann} equation
\begin{eqnarray}\label{21}
\frac{\partial \hat{\rho}(t)}{\partial t} = \frac{1}{i \hbar}
[\hat{H}_{MF}(t),\hat{\rho}(t)]  -  \gamma  \left(\hat{\rho}(t) -
\hat{\rho}^{(0)}\right) ,
\end{eqnarray}
where $\hat{\rho}^{(0)}$ is the equilibrium density matrix, and $\gamma$ is the
phenomenological relaxation rate. The equilibrium density matrix is diagonal in the basis
of eigenfunctions of the Hamiltonian (\ref{11})
\begin{equation}\label{22} (\hat{\rho}^{(0)})_{\nu,\nu'} =
\delta_{\nu,\nu'} f_\nu,
\end{equation}
where $f_\nu=[\exp(E_\nu/T)+1]^{-1}$ is the Fermi-Dirac distribution function. The
interaction with the external field is considered as a small perturbation and the density
matrix is sought as the series
\begin{equation}\label{23}
  \hat{\rho}=\hat{\rho}^{(0)}+\hat{\rho}^{(1)}+\hat{\rho}^{(2)}+\hat{\rho}^{(3)}
  +\ldots
\end{equation}
The terms in the series (\ref{23}) satisfy the recurrent equation
\begin{eqnarray}\label{24}
{(\hat{\rho}^{(i)}(t))_{\nu,\nu'}}  = \frac{1}{i \hbar} \int_{-\infty}^{t} dt' (
[\hat{H}_{int}(t'),\hat{\rho}^{(i-1)}(t')] )_{\nu,\nu'} \cr \times
 e^{(i \omega_{\nu,\nu'} + \gamma) (t' - t)},
\end{eqnarray}
where $\omega_{\nu,\nu'} =  (E_{\nu} - E_{\nu'}) /{\hbar}$.

Let the scalar potential in the graphene layers is equal to
$\varphi(\mathbf{r},t)=\varphi_0 \sin(q_x x-\omega t)$. It corresponds to the electric
field of the incident wave $\mathbf{E}=\mathbf{E}_{0}\cos(\mathbf{q}\mathbf{r}-\omega
t)$, where $\mathbf{E}_{0}=(E_{0x},0,E_{0z})$, $\mathbf{q}=(q_x,0,q_z)$, and $E_{0x}=-q_x
\varphi_0$.  Nonlinear in $\varphi_0$ part of electron density oscillations  is given by
the equation
\begin{equation}\label{72}
n^{(3)}_+(\mathbf{r},t) = \mathrm{Tr}[ \hat{\rho}^{(3)}(t) \hat{n}_+(\mathbf{r}) ].
\end{equation}
The 3rd harmonic term in the electron density oscillations reads
\begin{equation}\label{72a}
n^{(3\omega)}_{+}(\mathbf{r},t) = n^{(3\omega)}_{0}e^{3 i (q_x x-\omega t)}+c.c.,
\end{equation}
where the explicit expression for $n^{(3\omega)}_{0}$ can be obtained from Eq.
(\ref{72}).
 Using the continuity equation $  {e\partial n_+} /\partial t +\nabla
 \mathbf{j}_+=0$
one finds
\begin{equation}\label{72b}
j^{(3\omega)}_{x,+}(\mathbf{r},t) = \frac{e \omega n^{(3\omega)}_{0}}{q_x}e^{3 i (q_x
x-\omega t)}+c.c.
\end{equation}
Eqs. (\ref{72})-(\ref{72b}) yield the following expression for the current
\begin{eqnarray}\label{72c}
  j^{(3\omega)}_{x,+}(\mathbf{r},t)=e^{3i(q_x x-\omega t)}\sigma_{+++}^{(3)}(\mathbf{q}_x,\mathbf{q}_x,\mathbf{q}_x;
  \omega,\omega,\omega) E^3_{0x}\cr + c.c.,
\end{eqnarray}
where
\begin{eqnarray} \label{73}
\sigma^{(3)}_{+++}(\mathbf{q}_x,\mathbf{q}_x,\mathbf{q}_x; \omega,\omega,\omega) =
-i\frac{e^4 \omega}{2 S q_x^4}\cr \times \sum_{\nu_1,\nu_2,\nu_3,\nu_4}
\delta_{\mathbf{k}_2,\mathbf{k}_1+\mathbf{q}_x}
\delta_{\mathbf{k}_3,\mathbf{k}_2+\mathbf{q}_x}
\delta_{\mathbf{k}_4,\mathbf{k}_3+\mathbf{q}_x}
     R_{12}  R_{23}  R_{34}  R_{41}
    \cr \times
     \frac{1}
{ E_{1} - E_{4} - 3 \hbar \omega - i\hbar\gamma } \bigg[ \frac{ 1 }{
E_{1} - E_{3} - 2\hbar \omega - i\hbar\gamma }\cr\times \bigg(
\frac{ f_{1} - f_{2} }{ E_{1} - E_{2} - \hbar \omega - i\hbar\gamma
}  - \frac{ f_{2} - f_{3} }{ E_{2} - E_{3} - \hbar \omega -
i\hbar\gamma } \bigg) \cr - \frac{ 1 }{ E_{2} - E_{4} - 2\hbar
\omega - i\hbar\gamma }\cr \times \bigg( \frac{ f_{2} - f_{3} }{
E_{2} - E_{3} - \hbar \omega - i\hbar\gamma }  - \frac{ f_{3} -
f_{4} }{ E_{3} - E_{4} - \hbar \omega - i\hbar\gamma } \bigg) \bigg]
\end{eqnarray}
is the 3rd order nonlinear conductivity.  In Eq. (\ref{73}) the shorthand notations
$E_i\equiv E_{\nu_i}$, $f_i \equiv f_{\nu_i}$, and $R_{ik}\equiv R_{\nu_i,\nu_k}$ are
used. The factor 4 that accounts the sum over 4 Dirac components is included in
(\ref{73}). Taking the limit $q_x\to 0$ we arrive at the relations (\ref{2}) and
(\ref{4}) with $\sigma_{+++}^{(3)}(\omega,\omega,\omega)=\lim_{q_x\to 0}
\sigma^{(3)}_{+++}(\mathbf{q}_x,\mathbf{q}_x,\mathbf{q}_x; \omega,\omega,\omega)$ .

\section{Results and discussion}

 {It was shown in \cite{21a} that in a DMLG system the order parameter
$\Delta_{\mathbf{k},\lambda}$  is peaked at the Fermi surface and decreases far from this
surface. In DBLG system a regime with almost constant $\Delta_{\mathbf{k},\lambda}$ at
$k<4 k_F$ was found in \cite{27}.   In our computations of THG intensity  we neglect the
wave vector dependence of the order parameter and replace $\Delta_{k_F,\lambda}$ with
$\Delta=E_g/2$, where $E_g$ is the energy gap. Such an approximation works well near the
Fermi surface.  The resonant features in the THG intensity described below  are caused,
in the main part, by the transitions between the electron states near the Fermi surface.
It justifies the use of the approximation $\Delta_{k_F,\lambda}=\Delta$.}

\begin{figure}[h]
     \center{\includegraphics[width=0.4\textwidth]{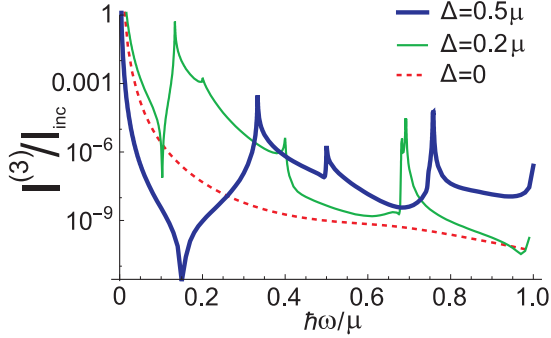}}
     \caption{THG intensity for the double monolayer graphene in the state with the electron-hole
     pairing ($\Delta=0.5\mu,0.2\mu$) and in the unpaired state ($\Delta=0$)}\label{f1}
     \end{figure}

\begin{figure}[h]
     \center{\includegraphics[width=0.4\textwidth]{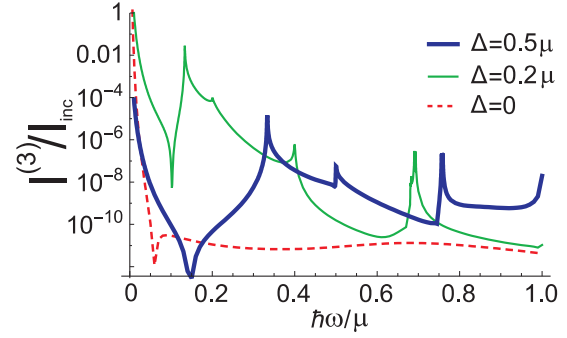}}
     \caption{The same as in Fig. \ref{f1} for the double bilayer graphene}\label{f2}
     \end{figure}

We  fix the temperature and the relaxation rate as $T=0.1 \mu$ and $\hbar\gamma=
0.001\mu$. The ratio of the THG intensity  to the incident wave intensity versus the
incident wave frequency is shown in Figs. \ref{f1} and \ref{f2}. Fig. \ref{f1}
corresponds to the DMLG system and Fig. \ref{f2}, to the DBLG system. The dependencies
are presented for the paired state for two different values of the order parameter and
for the unpaired state ($\Delta=0$). The absolute value of the ratio $I^{(3)}/I_{inc}$ in
Figs. \ref{f1} and \ref{f2} is computed for $\mu=0.01$ eV and $I_{inc}=5$ W/cm$^2$.  The
dependencies presented are scaled as
$$\frac{I^{(3)}}{I_{inc}}\propto \frac{I_{inc}^2}{\mu^8}$$ for the DMLG system,
and $$\frac{I^{(3)}}{I_{inc}}\propto \frac{I_{inc}^2}{\mu^6}$$ for the DBLG system.

 {One can see that the dependencies obtained have a number of peaks. There are
three peaks that correspond to the incident photon energies $\hbar\omega=(2/3){\Delta}$,
$\hbar\omega=\Delta$, and $\hbar\omega=2\Delta$. Numerically the peaks appear because
denominators in the expression for the nonlinear conductivity (\ref{73}) become resonant
when the incident photon energy $\hbar\omega$ exceeds $E_g/3$, $E_g/2$ and $E_g$.
Physically, it means opening of new channels of nonlinear absorption.} In addition a
double peak emerges at $\hbar\omega\approx(2/3)\sqrt{\mu^2+\Delta^2}$.

 {The peaks in THG intensity are rather sharp. We connect it with the divergence
of the electron density of states in the paired state. Indeed, in the normal state the
sum of the density of states in the electron and the hole layers is the constant
$\nu_e+\nu_h=2\nu_F$ in the interval $-\mu\leq \varepsilon\leq \mu$, where $\nu_F$ is the
density of states at the Fermi level for a isolated monolayer (bilayer) graphene, and the
energy $\varepsilon$ is counted from the Fermi level.  For the paired state simple
calculations yield
 $\nu(\varepsilon)=2 \nu_F\varepsilon/\sqrt{\varepsilon^2-\Delta^2}$ (in this case  the density of
states cannot be separated to the electron and the hole parts). This function diverges at
$\varepsilon=\pm \Delta$. The influence of pairing on the spectrum and on the density of
states in a DMLG system is illustrated in Fig. \ref{f3}. Note that far from the Fermi
level the energy spectra for the paired and the normal states approach each other. It
means that  the wave vector dependence of $\Delta_{\mathbf{k},\lambda}$ yields only an
inessential correction of the spectrum that can be considered as another justification of
the approximation $\Delta_{k_F,\lambda}=\Delta$.}

 { \begin{figure}[h]
     \center{\includegraphics[width=0.4\textwidth]{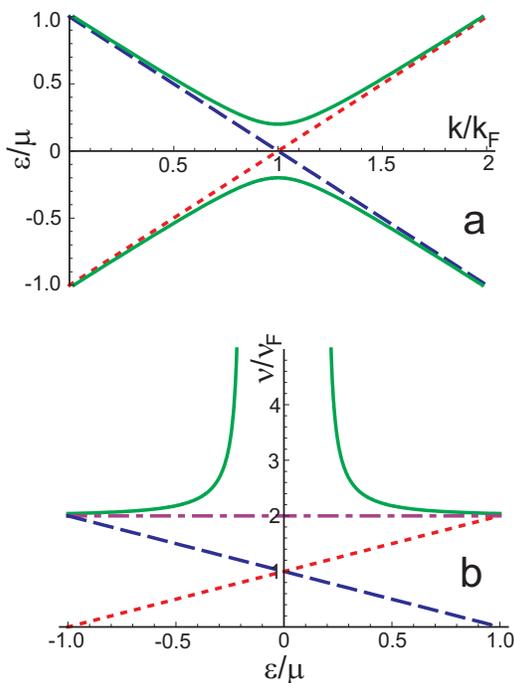}}
     \caption{ {Energy spectrum (a) and density of states (b) near the Fermi
     level in a double monolayer graphene for the paired state with  $\Delta=0.2\mu$ (solid lines)
     and for the normal state (dotted lines, dashed and dash-dotted lines). Dotted lines
     correspond to the electron-doped graphene layer, and dashed lines, to the hole-doped one.
     The total (electron plus hole) density of states is shown by the dash-dotted line.
     The energy difference between the Dirac points in the electron and hole layers is equal
     $2\mu$. The
     energy $\varepsilon$ is counted from
     the Fermi level, and the wave vector $k$ is counted from the Dirac point.}}\label{f3}
     \end{figure}}

The dependencies that correspond to $\Delta=0$ do not demonstrate any peaks.
 {The peaks predicted in \cite{3,4,6} emerges at lower temperature ($T\lesssim
10^{-2}\mu$). Under transition to the normal state the double peak  at
$\hbar\omega\approx(2/3)\sqrt{\mu^2+\Delta^2}$ is transformed to the peak at
$\hbar\omega=(2/3)\mu$, and the peaks at $\hbar\omega=(2/3){\Delta}$,
$\hbar\omega=\Delta$, and $\hbar\omega=2\Delta$ disappear.}

One can see that the electron-hole pairing   {causes} a huge increase of the intensity of
THG  {in a certain frequency range.} At the main peak $\hbar\omega=(2/3)\Delta$ the THG
intensity is in 8 orders of magnitude greater than one for the unpaired state. For the
out-of-resonance frequencies the factor of THG enhancement is also very large (about
$10^3$).  {We note that the enhancement of THG is not observed at large ($\hbar\omega\gg
\Delta$) frequencies and at small ($\hbar\omega\ll \Delta$). In the latter case the
pairing even suppresses the THG. It can be understood from the classical picture of a
tightly bound electron-hole pair that does not respond to a static electric field if the
same field is applied to the electron and the hole component of the pair.

Thus we consider that the enhancement of THG intensity is physically caused by the
appearance of new resonant frequencies connected with the gap $E_g=2\Delta$, and by the
divergence of the  density of states near the gap.

From the practical point of view} the enhancement means that strong nonlinear response in
the double layer graphene with electron-hole pairing can be observed at much smaller
incident wave intensity than in the system where the pairing does not occur.

In conclusion, we have shown that the electron-hole pairing in the double layer graphene
system results in  the strong enhancement of the nonlinear response to the
electromagnetic radiation. We predict the appearance of  a number of peaks in the THG
intensity. The main peak corresponds to the frequency equals to one third of the energy
gap in the spectrum and the intensity of this peak exceeds  in many orders the THG
intensity at the same frequency in the unpaired state. The impact of the pairing is
basically the same for the double monolayer and double bilayer graphene systems and we
expect THG enhancement in other systems  in which the electron-hole pairing may occur.

\acknowledgments This work was supported by the State Fund for Fundamental Research of
Ukraine, project No 33683.


\end{document}